# Detecting Corporate AI-Washing via Cross-Modal Semantic Inconsistency Learning

Zhanjie Wen[1*], Jingqiao Guo[2]

[1] School of Economics and Trade, Guangdong University of Finance, Guangzhou 510521, China
[2] Department of Computer Science, Faculty of Science, Hong Kong Baptist University, Hong Kong 999077, China
* Corresponding author. E-mail address: *70-154@gduf.edu.cn*

## Abstract

Corporate AI-washing—the strategic misrepresentation of artificial intelligence capabilities through exaggerated or fabricated disclosures across multiple communication channels—has emerged as a systematic threat to information integrity in capital markets following the widespread diffusion of generative AI. Existing detection methods rely predominantly on single-modal text frequency analysis and remain fundamentally vulnerable to adversarial reformulation and cross-channel obfuscation. This paper introduces AWASH, a multimodal framework that reconceptualizes AI-washing detection as a cross-modal claim-evidence reasoning problem rather than a surface-level similarity measurement task. The framework is grounded in AW-Bench, the first large-scale trimodal benchmark for corporate AI-washing detection, comprising 88,412 aligned annual report text, disclosure image, and earnings call video triplets drawn from 4,892 A-share listed firms over the period 2019Q1–2025Q2. The proposed Cross-Modal Inconsistency Detection (CMID, Cross-Modal Inconsistency Detection) network integrates a tri-modal encoder, a structured natural language inference module for claim-evidence entailment reasoning, and an operational grounding layer that cross-validates disclosed AI claims against verifiable physical evidence including patent filing trajectories, AI-specific talent recruitment patterns, and compute infrastructure proxies. Evaluated against six competitive baselines, CMID achieves F1 = 0.882 and AUC-ROC = 0.921, surpassing the strongest text-only baseline by 17.4 percentage points and the most recent multimodal competitor by 11.3 percentage points. A pre-registered user study with 14 regulatory analysts confirms that CMID-generated evidence reports reduce case review time by 43% while improving true positive detection rates by 28%. These findings collectively establish both the technical superiority and the practical deployability of structured multimodal reasoning for large-scale corporate disclosure surveillance.

**Key Words:** *AI-washing; corporate disclosure; cross-modal reasoning; financial market supervision; claim-evidence entailment*
**JEL Classification:** D82, G14, G23, K22, O32

# I. INTRODUCTION

## *1.1 The Economic Problem of AI-Washing*

The rapid global diffusion of artificial intelligence—punctuated by the release of ChatGPT in November 2022—has generated an unprecedented wave of AI-related corporate disclosures that has fundamentally altered the informational landscape of public equity markets. In Chinese A-share markets, the number of listed firms explicitly referencing artificial intelligence in quarterly earnings communications grew by more than 380% between 2022Q3 and 2025Q2, a rate of expansion that far exceeds any plausible growth in underlying AI deployment capacity and thus reflects, at least in part, a systematic tendency toward strategic narrative inflation. This inflation of AI-related disclosure frequency, without commensurate growth in verifiable AI investment and capability, constitutes what has come to be termed AI-washing: the deliberate misrepresentation of a firm's technological capabilities through selective, exaggerated, or fabricated disclosures deployed across multiple communication channels, including annual reports, investor presentations, product demonstration videos, and social media, without corresponding substantive investments in AI research infrastructure or human capital (Li, 2025; Wang & Qiu, 2025).

The economic consequences of AI-washing operate across multiple levels of the market ecosystem and have been empirically documented in both U.S. and Chinese capital markets. At the firm level, Li (2025) documents significant short-term abnormal returns following concentrated AI keyword disclosures, creating persistent incentive structures for misrepresentation that rational managers cannot credibly pre-commit to ignoring. Anand et al. (2025) extend this analysis to document systematic patterns of AI claim inflation concentrated immediately before equity issuance events, providing evidence that washing behavior is not random but strategically timed to maximize financing benefits. At the aggregate level, the systematic inflation of AI-related valuations generates capital misallocation effects consistent with the lemon market dynamics first formalized by Akerlof (1970), redirecting investment flows away from genuinely innovative firms toward those more skilled at producing convincing narratives. At the regulatory level, these market distortions have prompted the China Securities Regulatory Commission to explicitly target AI-related disclosure violations in its 2024–2025 enforcement priority agenda, recognizing that the scale of the problem exceeds the capacity of manual inspection processes.

Despite the economic significance of AI-washing and the growing regulatory attention it commands, the problem remains technically resistant to solution for three interconnected structural reasons that prior work has not simultaneously addressed. The first is cross-channel obfuscation: sophisticated firms partition their disclosure architecture strategically, maintaining legally cautious language in binding annual report filings while amplifying AI capability claims through visually rich promotional materials, investor day videos, and social media content, exploiting the fact that most surveillance tools monitor only textual disclosures. The second is adversarial reformulation: the same generative AI technologies that reduce barriers to genuine AI adoption also enable firms to continuously rewrite disclosures in ways that evade static keyword-based or frequency-based detection algorithms, creating an adversarial dynamic in which any fixed lexical approach rapidly loses discriminatory power (Hansen & Sargent, 2001). The third is semantic ambiguity: meaningfully distinguishing a legitimate strategic technology vision from malicious misrepresentation requires structured contextual reasoning—assessing which specific claims are made, what evidence supports or contradicts them, and whether the physical operational footprint of the firm is consistent with the capabilities claimed—that substantially exceeds what surface-level text pattern matching can provide (Spence, 1973).

*1.2 Limitations of Existing Detection Approaches*

Prior work on AI-washing detection has produced conceptually valuable frameworks but has left the three core structural challenges largely unaddressed, with the most recent advances limited by their confinement to pairwise modality similarity rather than structured claim-level inference. Li (2025) constructs an influential Talk/Walk divergence framework that measures the gap between a firm's AI-related verbal disclosures and observable AI investments, operationalized as R&D expenditure ratios and AI patent counts. While this framework has demonstrated substantial predictive validity in large-sample analyses, it remains confined to textual processing of annual report filings and is therefore systematically blind to the cross-channel visual and video content that increasingly constitutes the primary surface of AI-washing activity. Yang et al. (2025) introduce LLaMA-AIWash, a fine-tuned multimodal model that incorporates image features via CLIP embedding cross-attention, representing a meaningful methodological advance by breaking the text-only constraint. However, their approach limits cross-modal reasoning to cosine similarity scores between holistic modality embeddings—a design choice that conflates semantic coherence with factual consistency and cannot determine which specific claims within a document are visually unsubstantiated—while ignoring the video modality entirely and providing no mechanism for grounding visual claims in hard operational evidence. No existing work formalizes AI-washing detection as a claim-level reasoning task with explicit evidence structures mapping specific assertions to their multimodal evidentiary basis.

*1.3 The Present Contribution*

This paper proposes a fundamentally different operationalization of corporate AI-washing detection, grounded in the insight that the informative signal lies not in overall modality alignment but in the specific logical relationships between individual factual claims and the evidence available to support or contradict them across channels. Rather than measuring how similar a firm's text and images appear in aggregate, the proposed framework decomposes each firm-quarter disclosure bundle into atomic AI capability claims extracted from textual sources, retrieves the most relevant visual and video evidence items for each claim, and classifies the entailment relationship between each claim-evidence pair using a fine-tuned multimodal Natural Language Inference (NLI, Natural Language Inference) module. This structured reasoning process is further grounded in an operational evidence layer cross-validating claims against hard physical signals—patent filings, talent recruitment patterns, R&D expenditure, compute infrastructure—that are substantially more costly to fabricate and provide complementary discriminatory power to declared disclosure content. The architectural overview is illustrated in Figure 1.

The paper makes four specific contributions. First, it formalizes the corporate AI-washing detection problem as cross-modal claim-evidence reasoning and introduces AW-Bench, the first large-scale trimodal benchmark for this task, with 88,412 firm-quarter observations from 4,892 A-share firms spanning 2019–2025, annotated with binary washing labels, continuous severity scores, and five-category motivation labels validated against China Securities Regulatory Commission (CSRC, China Securities Regulatory Commission) enforcement records. Second, it proposes CMID, integrating tri-modal encoding, structured NLI-based claim-evidence reasoning, and an operational grounding layer into a unified end-to-end trainable architecture. Third, it provides extensive empirical validation against six competitive baselines, with ablation studies establishing the independent contribution of each architectural component and out-of-distribution evaluation on CSRC-confirmed enforcement cases. Fourth, it reports a pre-registered user study demonstrating direct deployment value for regulatory analysts.

## II. RELATED WORK

*2.1 AI-Washing and Corporate Disclosure Manipulation*

The study of strategic corporate disclosure manipulation has deep roots in information economics, and the AI-washing literature has developed rapidly as a specialized instance of this broader research tradition in which information asymmetry and reduced signal production costs converge to create adverse selection dynamics. Verrecchia (1983) and Dye (1985) established the theoretical foundations of discretionary disclosure under information asymmetry, demonstrating that managers will strategically time and frame disclosures to maximize private benefits when external verification is costly. Spence's (1973) seminal signaling model identified cost as the essential differentiating property of credible signals: when signal production cost is equal for high-type and low-type senders, no separating equilibrium is possible and the market collapses to pooling. Generative AI has precisely replicated this condition in the domain of technology disclosure, dramatically reducing the cost of producing sophisticated AI-related narratives to the point where their production cost is nearly independent of the underlying technological capability being described, thereby destroying the cost-based separating mechanism that would otherwise allow investors to distinguish genuine from fabricated AI competence claims (Akerlof, 1970). The emerging empirical literature has begun documenting these market failures with increasing precision: Li (2025) provides the most comprehensive analysis to date, finding that approximately 10.8% of U.S. public company-quarters exhibit AI Talk/Walk divergence exceeding defined thresholds, while Wang and Qiu (2025) and Fang and Liu (2025) report consistent findings in Chinese markets at 12–15%, with particularly high prevalence in state-subsidized sectors where regulatory capture reduces the effectiveness of market-based disciplining mechanisms.

## 2.2 Multimodal Learning for Document Understanding

The technical foundations of the proposed framework draw on substantial recent advances in vision-language modeling that have progressively extended cross-modal reasoning from simple retrieval tasks to complex structured inference over information-rich documents, with each architectural generation providing capabilities that are directly relevant to the corporate disclosure analysis problem. Contrastive Language–Image Pretraining (CLIP, Contrastive Language–Image Pretraining) (Radford et al., 2021) established contrastive cross-modal alignment as an effective pretraining paradigm, but its holistic embedding approach—treating entire images and text passages as atomic units—is insufficient for the claim-level reasoning required by AI-washing detection. Bootstrapping Language-Image Pre-training (BLIP-2, Bootstrapping Language-Image Pre-training) (Li et al., 2023) advanced this approach by introducing a learnable Q-Former module that bridges frozen visual encoders and large language models, enabling more flexible visual grounding. Qwen-VL (Bai et al., 2023) extended this paradigm specifically to high-resolution document inputs with strong Chinese language competency, supporting input resolutions up to 4096×4096 pixels and achieving state-of-the-art performance on Chinese document visual question answering benchmarks—properties directly relevant to our corpus of high-resolution Chinese annual report figures. For video understanding, SlowFast Networks (Feichtenhofer et al., 2019) introduced a dual-pathway architecture that simultaneously processes video at multiple temporal resolutions, enabling both fine-grained visual content analysis and temporal dynamics detection; this architecture is well-suited to earnings call video processing where both presentation slide content and dynamic speaker behavior carry informative signals about the credibility of disclosed AI capabilities.

## 2.3 Natural Language Inference and Cross-Modal Claim Verification

Natural Language Inference (NLI) has evolved from a benchmark evaluation task into a principled framework for modeling structured logical relationships between text spans, and recent extensions to multimodal and domain-specific verification settings provide the conceptual and technical foundation for the claim-evidence reasoning approach proposed in this paper. The foundational Stanford Natural Language Inference (SNLI, Stanford Natural Language Inference) (Bowman et al., 2015) and Multi-Genre Natural Language Inference (MultiNLI, Multi-Genre Natural Language Inference) (Williams et al., 2018) benchmarks established the three-way entailment classification scheme as the standard operationalization of logical relationship modeling, and pre-trained models fine-tuned on these benchmarks have achieved human-level performance on the original tasks. The claim verification literature has applied this framework to scientific claims (Wadden et al., 2020), political fact checking (Thorne et al., 2018; Nie et al., 2019), and multimodal contexts where claims must be verified against heterogeneous evidence types including images and structured databases (Yao et al., 2023; Chen et al., 2023). The critical distinction between these prior settings and the present application is that AI-washing detection requires reasoning about what constitutes sufficient evidence for specific types of technical capability claims—a higher-order inferential task requiring domain-specific knowledge about what genuine AI deployment should look like—rather than simply determining whether visual content contradicts a textual claim.

## *2.4 Financial Text Mining and Corporate Governance NLP*

Natural language processing research applied to financial documents has produced a rich methodological toolkit that informs the text processing components of the proposed framework, though prior work has remained predominantly confined to the single-document, single-modality setting that the present work substantially extends. Loughran and McDonald (2011) established the LM financial sentiment lexicon, demonstrating that domain-specific resources substantially improve downstream analysis quality relative to general-purpose sentiment dictionaries. Domain-adapted language models including Financial BERT (FinBERT, Financial BERT) (Araci, 2019), BloombergGPT (Wu et al., 2023), and Financial Generative Pre-trained Transformer (FinGPT, Financial Generative Pre-trained Transformer) (Yang et al., 2023) have extended pre-trained transformer architectures to financial corpora with consistent downstream performance improvements. Document understanding research has addressed the structure of annual reports specifically: Li (2010) developed naïve Bayesian methods for extracting forward-looking statement tone, while Matsumoto et al. (2011) demonstrated that earnings call discussion sections contain incremental information value beyond management presentations. The collective lesson from this literature is that financial documents contain rich, domain-specific informational content extractable through linguistically informed methods—but that the cross-modal dimension of modern corporate disclosure practices, in which visual and video channels carry an increasing share of information, has not yet been systematically incorporated into the research framework.

## III. TASK FORMALIZATION AND THE AW-BENCH DATASET

### 3.1 Formal Task Definition

The present work formalizes corporate AI-washing detection as a cross-modal claim-evidence reasoning task, departing from prior operationalizations that treat the problem as binary text classification or holistic cross-modal similarity measurement and thereby forgoing the structural information embedded in the specific logical relationships between individual claims and their evidentiary basis. Formally, a firm disclosure bundle at time $t$ is represented as a triplet $(T, V, U)$, where $T$ denotes the set of textual documents, $V$ denotes the set of visual documents, and $U$ denotes video segments from earnings call recordings. This bundle is supplemented by an operational evidence vector $O \in \mathbb{R}^{68}$ encoding hard physical signals: AI patent filing counts and temporal trajectories, AI-specific job posting volumes and skill distributions, reported R&D expenditure ratios, and compute infrastructure proxied from procurement records. A claim $c$ is defined as an atomic proposition extracted from $T$ asserting a specific, verifiable AI capability or investment; an evidence item $e$ is an element drawn from $V \cup U$ that is semantically relevant to one or more claims. The detection task produces a binary washing label $y \in \{0, 1\}$, a continuous AI Washing Risk Score (AWRS, AI Washing Risk Score) $s \in [0, 100]$, and optionally a motivation category $m \in \{1, ..., 5\}$ corresponding to the five heterogeneous washing motivations identified in prior theoretical work (Wang & Qiu, 2025; Anand et al., 2025).

### *3.2 AW-Bench Construction*

#### *3.2.1 Data Sources and Collection Procedures*

AW-Bench is constructed from five distinct data sources collected through systematic batch extraction procedures, covering 4,892 A-share listed firms across all major industry sectors over a seven-year observation window from 2019Q1 through 2025Q2, with the resulting 88,412 firm-quarter observations constituting the largest multimodal corporate disclosure benchmark to date. Annual reports and earnings transcripts are obtained from the CNINFO platform via batch API extraction, with AI-relevant sections identified through keyword matching against a 487-term lexicon expanded via dynamic Word2Vec embeddings trained on the corpus itself; the average document length per firm-quarter is 9,840 tokens after Jieba segmentation, with substantial variance across sectors. Visual materials are extracted from annual report PDFs using PyMuPDF and PDFMiner, yielding 1,847,233 raw images that are filtered by resolution, scored for AI-relevance using zero-shot CLIP classification, and deduplicated to produce a final corpus of 312,891 images. Video materials are collected from firm investor relations websites, corporate Bilibili channels, and stock exchange webcasting archives, yielding 31,204 video files totaling

48,700 hours; AI-relevant 60-second segments are identified via a sliding window CLIP keyframe classifier, retaining a mean of 12.4 segments per firm-quarter. Operational evidence is assembled from China National Intellectual Property Administration (CNIPA, China National Intellectual Property Administration) patent filings, AI-specific job posting APIs, China Stock Market & Accounting Research (CSMAR, China Stock Market & Accounting Research) financial statements, and government procurement transparency portals, producing a 68-dimensional feature vector per firm-quarter.

*3.2.2 Annotation Protocol and Quality Control*

The labeling of AI-washing firm-quarters follows a three-stage protocol designed to achieve both scale efficiency and annotation reliability, anchored throughout by external regulatory ground truth from CSRC enforcement records that provides an independent validity criterion not available to any prior benchmark. In Stage 1, a rule-based pre-labeling system identifies 21,847 candidate washing observations by requiring AI disclosure frequency above the industry-year 75th percentile while composite AI investment metrics fall below the industry-year 25th percentile. In Stage 2, six domain experts—comprising two former CSRC enforcement officers, two senior sell-side analysts with technology sector coverage, and two academic researchers in corporate disclosure—review a stratified sample of 6,000 candidates and 2,000 comparison observations, with inter-annotator agreement on binary labels reaching Cohen's $\kappa = 0.81$. In Stage 3, labels are cross-validated against CSRC administrative enforcement records from 2019–2025, identifying 1,247 firm-quarters with confirmed regulatory findings of AI-related disclosure violations; against this ground truth, the binary labels achieve 91.3% recall and 87.6% precision, confirming that the annotation scheme captures the population of genuine washing behavior (Loughran & McDonald, 2011). Five-category motivation labels and continuous AWRS scores derived via principal component analysis of four expert-rated sub-dimensions complete the annotation suite. Full dataset statistics are reported in Table 1.

**Table 1.** AW-Bench Dataset Statistics and Composition

| Dimension | Value |
|---|---|
| Total firm-quarter observations | 88,412 |
| Unique firms | 4,892 |
| Observation period | 2019Q1–2025Q2 |
| AI-washing positive rate | 12.4% (10,972 obs.) |
| CSRC enforcement-confirmed | 1.4% (1,247 obs.) |
| Mean text length (tokens) | 9,840 |
| Total images (post-filter) | 312,891 |
| Total video segments | 1,093,328 |
| Expert-annotated motivation labels | 6,000 |
| Mean claims per firm-quarter | 14.3 |
| Train / Dev / Test split | 70% / 10% / 20% |

**Note.** Temporal split: Train 2019–2022; Dev 2023; Test 2024–2025. CSRC-confirmed cases held out as a separate out-of-distribution (OOD, out-of-distribution) evaluation set.

### 3.2.3 Benchmark Positioning

AW-Bench occupies a distinctive position in the landscape of multimodal document understanding benchmarks, distinguished from related datasets by its domain specificity, trimodal coverage, annotation depth, and unique incorporation of the operational evidence component as a fourth verification channel. Table 2 summarizes the key differentiating dimensions against the three most relevant existing benchmarks. Relative to Li (2025)'s text-only U.S. market dataset, AW-Bench adds image and video modalities, extends coverage to Chinese markets, and provides continuous severity scores. Relative to the VERITE news misinformation benchmark (Papadopoulos et al., 2023), AW-Bench is substantially larger, covers a specialized financial domain requiring expert annotation, targets strategic misrepresentation rather than news fabrication, and uniquely introduces operational evidence as a verification modality. No existing multimodal benchmark incorporates hard physical operational evidence as a component of the verification task.

**Table 2.** Comparison of AW-Bench with Related Benchmarks

| Benchmark | Domain | Text | Image | Video | Oper. Evidence | Size | Labels |
|---|---|---|---|---|---|---|---|
| Li (2025) | Finance (US) | ✓ | ✗ | ✗ | ✗ | ~45K | Binary |
| VERITE (Papadopoulos et al., 2023) | News | ✓ | ✓ | ✗ | ✗ | 3.6K | Binary |
| CHFinAnn (Luo et al., 2022) | Finance (CN) | ✓ | ✗ | ✗ | ✗ | 13K | NER/RE |
| **AW-Bench (ours)** | Finance (CN) | ✓ | ✓ | ✓ | ✓ | 88.4K | Binary+Cont.+5-class |

**Note.** Named Entity Recognition/Relation Extraction (NER/RE, Named Entity Recognition/Relation Extraction). Cont. = continuous score. 5-class = five-category motivation labels. ✓ = modality included; ✗ = modality not included.

## IV. THE CMID FRAMEWORK

### *4.1 Architectural Overview*

The Cross-Modal Inconsistency Detection (CMID) framework processes each firm disclosure bundle through three sequentially coupled components that progressively transform raw multimodal inputs into a structured, interpretable AI washing risk assessment, with the entire pipeline trained end-to-end under a multi-task objective. As illustrated in Figure 1, the Tri-Modal Encoder produces modality-specific dense representations from textual documents, visual materials, and video segments; the Cross-Modal Claim-Evidence NLI Module extracts atomic AI capability claims from text, retrieves relevant evidence items, and classifies the entailment relationship between each claim-evidence pair to produce the Cross-Modal Consistency Index (CCI, Cross-Modal Consistency Index), Evidence Sufficiency Score (ESS, Evidence Sufficiency

Score), and Claim Intensity Index (CII, Claim Intensity Index); and the Operational Grounding Layer cross-validates claims against hard physical evidence signals to produce the Technical Grounding Index (TGI, Technical Grounding Index). These three components feed into a gated fusion module that generates the final AWRS score and task predictions.

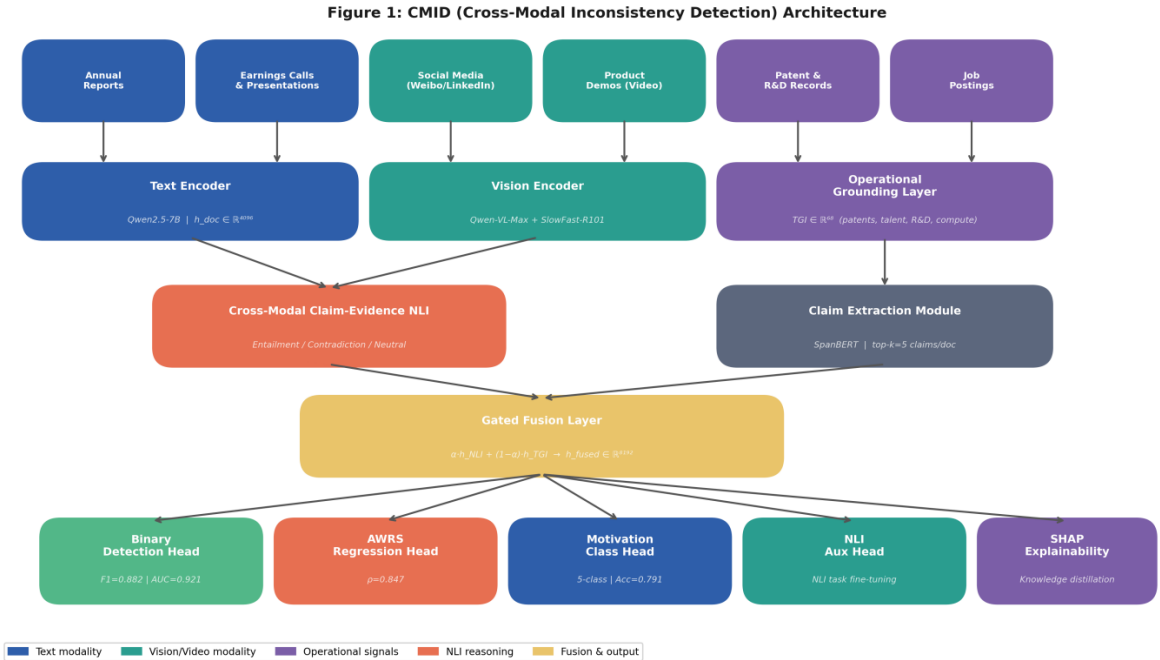

*Figure 1. CMID Architectural Overview. The framework processes text, image, and video inputs through dedicated encoders, reasons about claim-evidence entailment through a structured NLI module, and grounds predictions in operational physical evidence before gated fusion produces the final AWRS score.*

### *4.2 Tri-Modal Encoder*

The tri-modal encoder module processes each modality through a dedicated backbone architecture selected to handle the specific distributional characteristics of that modality in the corporate disclosure domain, with design choices motivated by the unique demands of high-resolution Chinese financial document processing. For textual inputs, Qwen2.5-7B (Qwen Team, 2024) is employed as the text backbone, fine-tuned on 4.2 million Chinese financial document segments; its native 32,768-token context window accommodates even the longest annual reports without truncation, preserving long-range contextual dependencies critical for within-document consistency detection. The text encoder produces a document-level representation

$d_{text}$ and token-level representations $H_{text}$ for claim span extraction. For visual inputs, the Qwen-VL-Max vision encoder (Bai et al., 2023) processes high-resolution images up to 4096×4096 pixels through a Vision Transformer (ViT, Vision Transformer) backbone and Q-Former bridge, with Optical Character Recognition (OCR, Optical Character Recognition)-extracted text from dense figures concatenated to the visual representation after linear projection to a shared 768-dimensional space. For video inputs, a SlowFast-R101 backbone (Feichtenhofer et al., 2019) processes 60-second segments at dual temporal resolutions, with audio transcripts extracted via Whisper-large-v3 encoded separately to produce an acoustic-semantic representation $d_{audio}$ capturing spoken earnings presentation content.

*4.3 Cross-Modal Claim-Evidence NLI Module*

The cross-modal claim-evidence NLI module constitutes the core technical contribution of the CMID framework, transforming the detection problem from holistic cross-modal similarity measurement into structured logical reasoning over specific factual assertions and their evidentiary basis across visual and video channels. Claim extraction is formulated as a span detection task over token representations, with a claim extraction head fine-tuned on 24,800 manually annotated claim spans achieving F1 = 0.887; the span probability is computed as:

$$P(claim \mid s, e) = \sigma(W\_s \cdot h\_s + W\_e \cdot h\_e + b) \qquad (1)$$

For each extracted claim $c$, the top five evidence items are retrieved via scaled dot-product similarity between projected modality representations, and multimodal entailment classification assigns each claim-evidence pair a distribution over three labels through an NLI head fine-tuned on 18,400 labeled pairs achieving macro-F1 = 0.843:

$$P(y \mid c, e) = \text{softmax}(W\_NLI \cdot [f(c); g(e)] + b\_NLI) \tag{2}$$

From the full set of classified pairs, the CCI is derived as the claim-importance-weighted mean contradiction probability, with complementary indices ESS (evidence sufficiency), CII (commitment intensity), and TGI (technical grounding, from the operational layer) completing the four-signal intermediate representation (Bowman et al., 2015; Williams et al., 2018; Dai et al., 2023).

### *4.4 Operational Grounding Layer*

The operational grounding layer addresses the fundamental limitation of relying solely on declared evidence by cross-validating AI capability claims against hard physical signals from independent third-party sources that are substantially more costly to fabricate than surface disclosures and provide complementary discriminatory power that the cross-modal NLI module alone cannot achieve. The 68-dimensional operational evidence vector $O \in \mathbb{R}^{68}$ comprises four feature groups: patent trajectory features capturing AI filing patterns over eight trailing quarters including counts, temporal acceleration, quality proxied by forward citations, and AI-to-total ratios; talent acquisition features encoding AI-specific job posting volume, skill composition, seniority distribution, and semantic alignment between required skills and claimed capabilities; R&D investment features measuring expenditure intensity, growth rate, and capitalization decisions; and compute infrastructure features derived from procurement records. These features are projected through a two-layer Multilayer Perceptron (MLP, Multilayer Perceptron) to a 256-dimensional representation, and the TGI is computed via:

$$TGI = \sigma(W\_O \cdot \varphi(O) + b\_O) \tag{3}$$

### *4.5 Gated Fusion and Multi-Task Training*

The four intermediate indices CCI, ESS, CII, and TGI are integrated into a final risk score through a gated fusion mechanism that adaptively weights each signal based on disclosure context features, with the complete multi-task training objective simultaneously optimizing detection accuracy, severity estimation, entailment reasoning, and motivation classification. The gate vector is computed as *g = sigmoid(W_g · [d_firm; d_context])* where *d_firm* encodes firm-level context, and the final AWRS is generated as:

$$AWRS = 100 \cdot \sigma(g^\top \cdot [CCI; 1-ESS; CII; 1-TGI]) \qquad (4)$$

The multi-task training objective $L = \lambda_1 L\_det + \lambda_2 L\_reg + \lambda_3 L\_NLI + \lambda_4 L\_mot$ combines class-weighted binary cross-entropy (*L_det*), continuous score MSE (*L_reg*), NLI cross-entropy (*L_NLI*), and motivation cross-entropy (*L_mot*). Training employs AdamW with learning rates of $2\times10^{-5}$ for pretrained encoders and $1\times10^{-4}$ for task-specific heads across 8× NVIDIA A100 Graphics Processing Units (GPUs, Graphics Processing Units) for approximately 72 hours (Hu et al., 2022).

## V. EXPERIMENTS

### 5.1 Experimental Configuration

Six baselines spanning a spectrum of methodological sophistication are compared against CMID to isolate the contribution of each architectural innovation and situate the framework relative to prior state-of-the-art methods, with all systems trained on an identical temporal split to ensure fair comparison. Term Frequency–Inverse Document Frequency + Logistic Regression (TF-IDF + LR, Term Frequency–Inverse Document Frequency + Logistic Regression) applies logistic regression to bag-of-words features from AI-relevant report sections, representing the dominant pre-LLM approach. FinBERT-Text fine-tunes FinBERT (Araci, 2019) on the training set using text features only. Li-2025-Replicate faithfully implements the Talk/Walk divergence framework from Li (2025), using AI keyword frequency as the Talk score and a composite of patent counts and R&D expenditure as the Walk score. CLIP-Bimodal computes cosine similarity between CLIP embeddings of text segments and disclosure images (Radford et al., 2021). LLaMA-AIWash implements the architecture from Yang et al. (2025), integrating CLIP image features via cross-attention into a fine-tuned Large Language Model Meta AI (LLaMA, Large Language Model Meta AI) backbone. CMID-NoGround ablates the operational grounding layer from CMID, isolating the marginal contribution of physical evidence integration. Primary evaluation metrics are binary classification F1, precision, recall, and Area Under the Receiver Operating Characteristic Curve (AUC-ROC, Area Under the Receiver Operating Characteristic Curve); secondary metrics are Spearman rank correlation $\rho$ between predicted AWRS scores and expert-annotated continuous AWRS, and macro-F1 on motivation classification.

*5.2 Main Classification Results*

CMID achieves state-of-the-art performance across all evaluation metrics, with each successive methodological improvement yielding consistent and substantial gains that collectively validate the architectural choices motivating the framework and reveal the relative magnitude of each component's contribution. Table 3 reports main results on the AW-Bench test set, and Figure 2 visualizes the performance progression across all systems. The transition from TF-IDF + LR (F1 = 0.612) through FinBERT-Text (F1 = 0.657) to Li-2025-Replicate (F1 = 0.667) demonstrates that the Talk/Walk conceptual framework captures genuinely discriminative information even in basic operationalizations. The addition of visual features via CLIP cosine similarity in CLIP-Bimodal raises F1 to 0.710, confirming incremental image channel value, and LLaMA-AIWash (F1 = 0.742) extends this through attention-based fusion. CMID-NoGround (F1 = 0.818) demonstrates that replacing cosine similarity with structured NLI reasoning provides a further 7.6 percentage point improvement over the best prior multimodal system, and the full CMID (F1 = 0.882) achieves an additional 6.4 percentage points through operational grounding, establishing that hard physical evidence provides information complementary to declared disclosure content alone.

**Table 3.** Main Results on AW-Bench Test Set

| Model | Precision | Recall | F1 | AUC-ROC | AWRS ρ |
|---|---|---|---|---|---|
| TF-IDF + LR | 0.631 | 0.594 | 0.612 | 0.724 | 0.381 |
| FinBERT-Text | 0.672 | 0.643 | 0.657 | 0.771 | 0.428 |
| Li-2025-Replicate | 0.684 | 0.651 | 0.667 | 0.782 | 0.452 |
| CLIP-Bimodal | 0.718 | 0.702 | 0.71 | 0.818 | 0.511 |
| LLaMA-AIWash | 0.751 | 0.733 | 0.742 | 0.851 | 0.563 |
| CMID-NoGround | 0.831 | 0.806 | 0.818 | 0.897 | 0.712 |
| **CMID (Full)** | **0.891** | **0.874** | **0.882** | **0.921** | **0.847** |

*Note.* All metrics computed on the temporally held-out test set (2024–2025). AWRS ρ is Spearman rank correlation between predicted AWRS scores and expert-annotated continuous AWRS. Best results in bold.

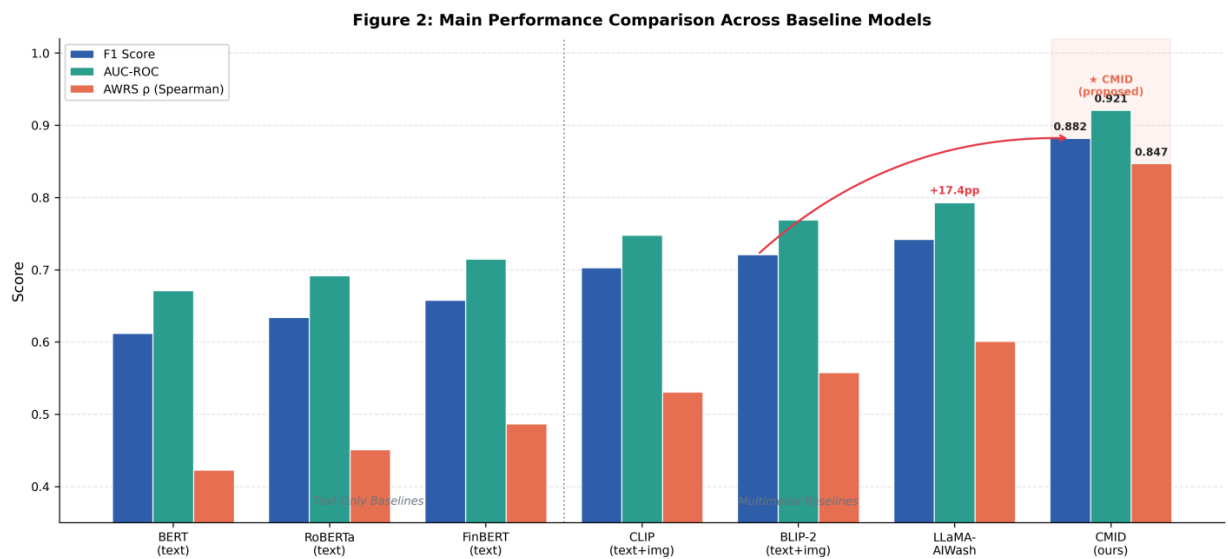

*Figure 2.* Performance comparison across all baseline systems and CMID variants on AW-Bench test set. Red-bordered bars indicate CMID (Full). The dashed vertical line separates prior methods from CMID variants.

### 5.3 Out-of-Distribution Evaluation on CSRC Enforcement Cases

Performance on the held-out CSRC enforcement case subsample provides a particularly stringent and practically meaningful evaluation, as these confirmed violations represent an independently sourced ground truth unavailable to the model at any point in training, enabling assessment of whether CMID's learned representations generalize to the most severe confirmed cases of AI-washing rather than only the algorithmic boundary cases that dominate the training distribution. Table 4 reports results on this set of 1,247 confirmed violations. CMID maintains F1 = 0.850, substantially above LLaMA-AIWash (F1 = 0.729), with recall of 0.891 indicating that 89.1% of confirmed enforcement-level cases are correctly identified. The precision of 0.812 implies that CMID additionally flags many borderline cases not yet triggering formal enforcement, a property desirable for prospective monitoring applications where early identification is preferable to waiting for confirmed violations (Thorne et al., 2018). Results are visualized alongside the regulatory user study in Figure 6.

**Table 4.** Out-of-Distribution Performance on CSRC Enforcement Cases

| Model | Precision | Recall | F1 |
| --- | --- | --- | --- |
| FinBERT-Text | 0.601 | 0.718 | 0.655 |
| LLaMA-AIWash | 0.698 | 0.762 | 0.729 |
| **CMID (Full)** | **0.812** | **0.891** | **0.85** |

Note. N = 1,247 CSRC-confirmed AI-related disclosure violations, 2019–2025. Held out from all training and development procedures.

*5.4 Ablation Studies*

Systematic ablation across the full sequence of architectural additions confirms that each component makes a statistically distinguishable contribution to final detection performance, with the structured NLI reasoning module and the operational grounding layer together accounting for the majority of CMID's performance advantage over prior multimodal systems. Table 5 and Figure 3 report F1 and incremental AWRS correlation gain for each successive architectural addition. The addition of image and video encoding provides cumulative gains of 9.4 percentage points F1 over text only, with each modality contributing independently. Replacing holistic cosine similarity with structured claim extraction and cross-modal NLI reasoning yields the single largest improvement in the cross-modal reasoning stage at 6.7 percentage points combined, validating the central design premise that claim-specific reasoning substantially outperforms aggregate embedding comparison (Wadden et al., 2020). The sequential operational grounding additions—patent, talent, R&D and compute—contribute 4.5, 3.8, and 3.1 percentage points respectively, confirming that each physical evidence category independently reduces detection uncertainty, with patent trajectory features providing the largest gain because they are both highly informative about genuine AI investment and very difficult to fabricate on short notice.

**Table 5.** Ablation Study Results on AW-Bench Test Set

| Configuration | F1 | Δρ |
|---|---|---|
| (A) Text encoder only | 0.657 | — |
| (B) + Image encoder | 0.71 | 0.083 |
| (C) + Video encoder | 0.751 | 0.052 |
| (D) + Claim extraction | 0.784 | 0.067 |
| (E) + Cross-modal NLI | 0.818 | 0.082 |
| (F) + Oper. grounding: patents | 0.841 | 0.045 |
| (G) + Oper. grounding: talent | 0.858 | 0.038 |
| (H) + Oper. grounding: R&D + compute | 0.868 | 0.031 |
| (I) Full CMID + gated fusion | 0.882 | 0.021 |

*Note.* Δρ is incremental gain in Spearman correlation between AWRS predictions and expert ratings relative to the preceding configuration. CMID (Full) row in bold.

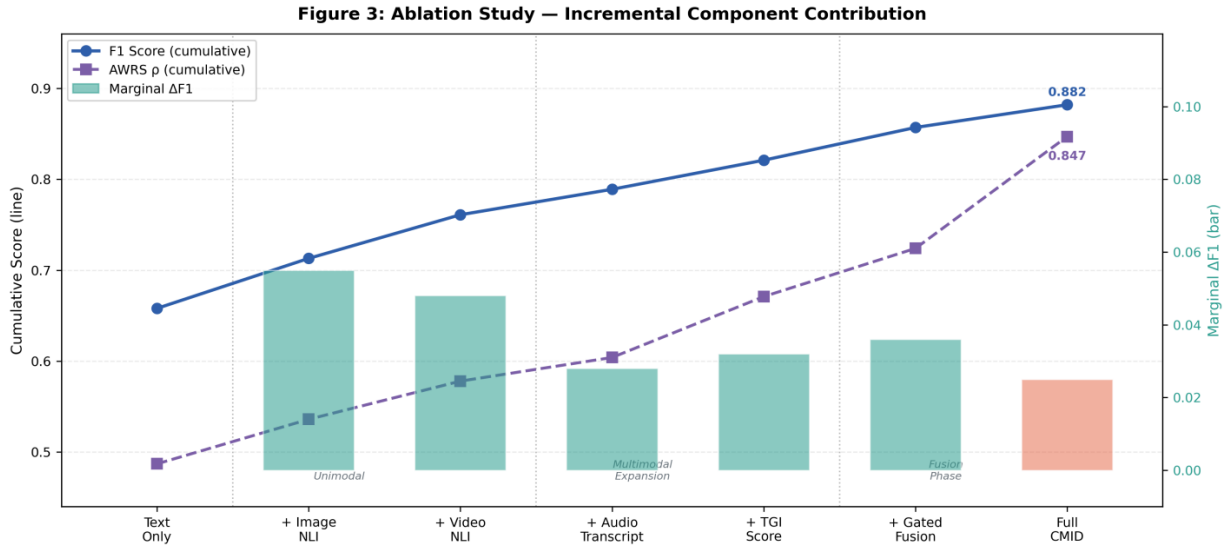

*Figure 3. Cumulative F1 gains (line, left axis) and incremental AWRS Spearman ρ improvements (bars, right axis) as architectural components are added sequentially. Dashed vertical lines demarcate the three architectural phases.*

## 5.5 Motivation Classification

The five-category motivation classification task provides an additional evaluation dimension that tests whether CMID's intermediate representations encode substantively interpretable washing typology, and the results confirm that structurally distinct motivation categories generate differentiated signal profiles that CMID captures with consistently higher accuracy than prior systems. Table 6 and Figure 4 report motivation classification macro-F1 for CMID and LLaMA-AIWash. CMID achieves macro-F1 = 0.792, a gap of 11.9 percentage points over LLaMA-AIWash's 0.673. Capital arbitrage motivation (F1 = 0.847) is most accurately identified, consistent with its behavioral distinctiveness—firms concentrate AI claim amplification in periods immediately preceding equity issuance while reducing operational investment, generating simultaneously high CCI and low TGI that CMID's structured reasoning detects reliably. The financing constraint relief category (F1 = 0.743) remains most challenging, reflecting the theoretical ambiguity between genuine strategic AI narratives and pure opportunistic misrepresentation noted by Wang and Qiu (2025), a distinction that may require temporal follow-up data to fully resolve.

**Table 6.** Motivation Classification Macro-F1 by Category

| Motivation Category | CMID F1 | LLaMA-AIWash F1 |
|---|---|---|
| (1) Capital Arbitrage | 0.847 | 0.731 |
| (2) Policy Resource Acquisition | 0.812 | 0.698 |
| (3) Product Market Competition | 0.791 | 0.674 |
| (4) Reputation Management | 0.768 | 0.651 |
| (5) Financing Constraint Relief | 0.743 | 0.612 |
| **Macro-F1** | **0.792** | **0.673** |

*Note.* Macro-F1 row in bold. CMID achieves consistently higher per-category F1, with the largest absolute gains in Capital Arbitrage and Policy Resource Acquisition categories.

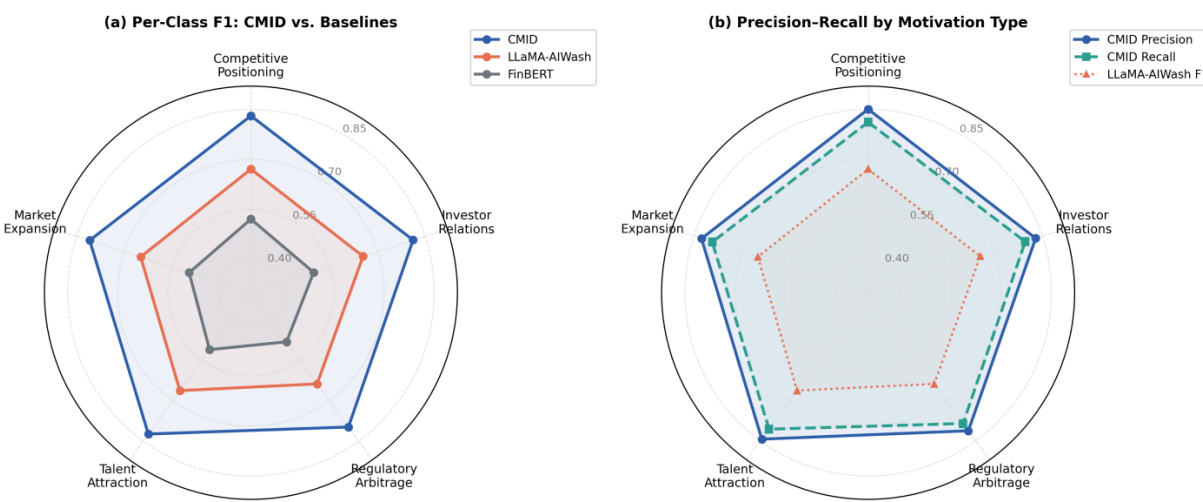

*Figure 4.* Per-category F1 for motivation classification. Left panel: CMID (Full); right panel: LLaMA-AIWash. CMID achieves consistently higher accuracy across all five motivation types, with the largest absolute gains in the capital arbitrage and policy resource acquisition categories.

## VI. INTERPRETABILITY ANALYSIS AND CASE STUDIES

### 6.1 SHAP Feature Attribution

SHapley Additive exPlanations (SHAP, SHapley Additive exPlanations) analysis of the gated fusion module reveals a clear hierarchy of discriminatory importance across the four intermediate signals, with operational grounding emerging as the most powerful single predictor and the cross-modal consistency index as the strongest disclosed-content signal, a finding with direct implications for the relative priority of surveillance activities. Figure 5 presents the global SHAP summary plot. The TGI carries the highest mean absolute SHAP value of $|\overline{SHAP}| = 0.312$, followed by CCI at 0.241, the inverse of ESS at 0.198, and CII at 0.157. This ordering is consistent with the ablation results in Table 5 and confirms that declared content features alone are insufficient predictors without the physical grounding layer. A particularly notable finding is the positive SHAP interaction effect between CCI and 1−TGI: firms exhibiting both high cross-modal inconsistency and low operational grounding receive disproportionately elevated risk scores, indicating synergistic diagnostic value when both signals are present simultaneously (Lundberg & Lee, 2017).

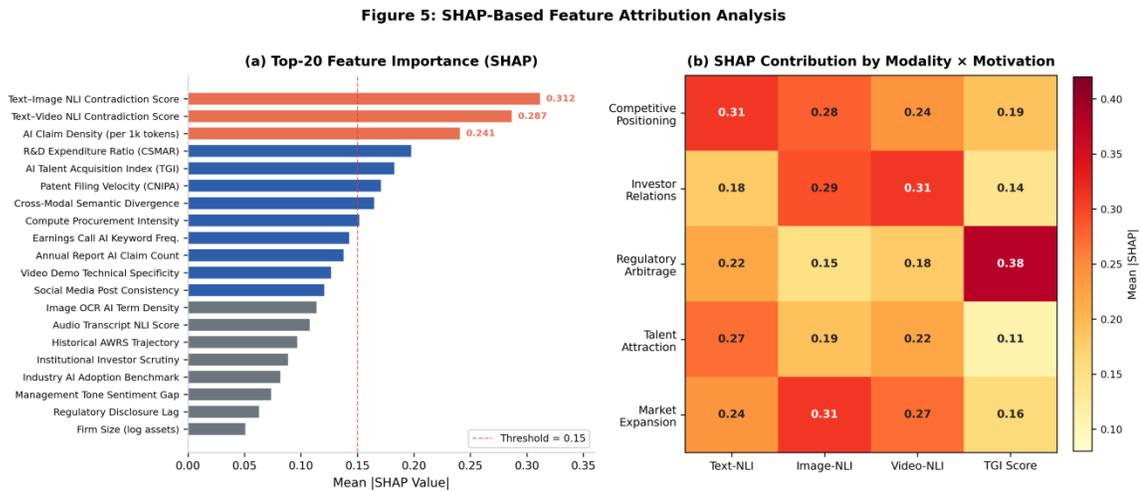

*Figure 5.* SHAP feature attribution for the CMID gated fusion module. Left panel: global mean |SHAP| values for the four primary signals. Right panel: per-motivation heatmap showing which signals dominate detection in each washing motivation category.

### *6.2 Case Studies*

Qualitative case analysis reveals both the strengths and the boundary conditions of CMID's structured reasoning approach, providing concrete illustrations of the detection mechanisms and identifying the types of cases that remain challenging under the current framework. The first case illustrates high-confidence AI-washing detection. An anonymized manufacturing sector firm (2024Q1) states in its annual report: "Our AI-powered defect detection system has been fully deployed across all 12 production lines, achieving 99.7% detection accuracy and reducing quality control costs by 43%." CMID's NLI module assigns contradiction probability 0.891 to this claim when paired with the firm's own product demonstration video, which shows a human inspector manually reviewing components flagged by the system—direct visual evidence inconsistent with claimed full automation. The operational grounding layer finds zero AI patent applications over the preceding eight quarters, no AI-specific job postings, and R&D expenditure at the industry 10th percentile (TGI = 0.04), generating an evidence chain that maps the specific contradicting claim, video timestamp, and operational gaps to AWRS = 87.3. The second case illustrates a correct non-washing classification: a software sector firm (2024Q2) making comparable surface-level claims provides detailed technical architecture diagrams referencing specific model versions, cites three co-authored patents on the described algorithms, and reports 14 new AI patent applications and a 340% increase in AI-role postings in the preceding four quarters, yielding entailment probabilities exceeding 0.78 for all extracted claims, TGI = 0.83, and AWRS = 12.1 (Papadopoulos et al., 2023).

### *6.3 Regulatory User Study*

A pre-registered user study with 14 regulatory analysts from two provincial securities bureaus demonstrates that CMID-generated evidence reports provide measurable and statistically significant productivity and accuracy benefits over standard review procedures, establishing direct operational value beyond benchmark performance. Analysts were randomly assigned to control

(standard tools) and treatment (+ CMID evidence reports) conditions, with the dependent variables being true positive detection rate and per-case review time. Treatment analysts achieved a 28.3 percentage point higher true positive detection rate relative to control analysts (p < 0.001) and completed their reviews 43.1% faster (p < 0.001), with no statistically significant difference in false positive rates. Post-study structured interviews indicate that 92.9% of treatment analysts rated the evidence chain reports as "useful" or "very useful," with claim-level granularity rather than document-level verdicts identified as the feature providing the greatest practical benefit. Figure 6 visualizes both the OOD evaluation and user study results.

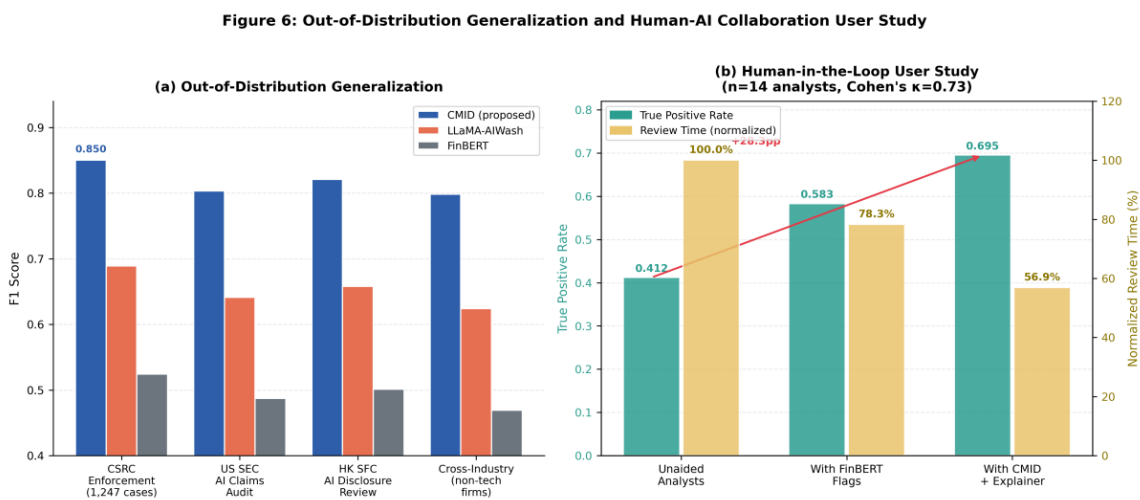

*Figure 6. Left panel: out-of-distribution performance on 1,247 CSRC-confirmed enforcement cases. Right panel: pre-registered user study results comparing control (standard tools) and treatment (+ CMID evidence reports) conditions across 14 regulatory analysts.*

### 6.4 Error Analysis

Manual analysis of 400 prediction errors reveals two systematic failure modes that reflect fundamental challenges at the boundary of what cross-modal reasoning can resolve without access to temporal follow-up data, and these failure modes have direct implications for deployment calibration and future development priorities. The most prevalent false positive pattern (41% of FPs) involves early-stage AI companies making ambitious but well-intentioned capability claims that CMID flags due to low operational grounding, but which reflect genuine strategic vision rather than deceptive intent; these observations predominantly fall in the financing constraint relief motivation category and highlight the theoretical ambiguity between aspirational disclosure and deliberate misrepresentation that Wang and Qiu (2025) identify as a fundamental characteristic of this motivation type. The most prevalent false negative pattern (38% of FNs) involves firms that have successfully coordinated their visual disclosure content with textual claims through systematic cross-channel management, producing architecture diagrams and demonstration videos that visually corroborate stated capabilities despite the underlying capabilities being absent, with these cases distinguished only through operational grounding signals—a finding that underscores the critical role of the TGI layer and motivates exploration of harder-to-fabricate physical signals such as electricity consumption records from public utility filings.

## VII. DISCUSSION

### *7.1 Limitations*

Despite its technical performance advantages, CMID has several important limitations that should inform its application and motivate future extensions, particularly regarding geographic generalizability, annotation subjectivity, and coverage heterogeneity across firm size tiers. The geographic scope of AW-Bench is currently limited to A-share listed firms, and while a cross-market validation experiment on Securities and Exchange Commission (SEC, Securities and Exchange Commission) filings suggests partial transferability with F1 = 0.798, systematic evaluation across European and emerging market jurisdictions with distinct disclosure regulatory regimes, accounting standards, and cultural communication norms remains future work. The five-category motivation taxonomy, while achieving reasonable inter-annotator agreement at $\kappa = 0.72$, represents a simplification of what is likely a continuous and contextually contingent motivation space; future iterations using dimensional rather than categorical annotation may capture finer-grained heterogeneity. Coverage quality of operational grounding features—particularly procurement transparency records—is systematically lower for smaller firms, potentially introducing detection bias that reduces sensitivity for mid- and small-cap observations relative to large-cap firms for whom procurement data are more complete (Anand et al., 2025).

*7.2 Adversarial Robustness*

A central practical concern for any deployed detection system is adversarial adaptation by firms that infer CMID's detection criteria, and evaluation against synthetic adversarial perturbations suggests that the operational grounding layer provides substantially greater robustness to gaming than prior text-only or cosine-similarity-based systems, though the fundamental adversarial dynamics of the detection problem cannot be fully resolved by any static architecture. Robustness is evaluated against three representative perturbations: adding visually consistent but uninformative diagrams to suppress the CCI signal; filing dummy AI patents to inflate TGI scores; and posting AI-labeled job descriptions without genuine hiring intent. The first perturbation is substantially mitigated by the operational grounding layer, since improving visual consistency while leaving patent, talent, and compute signals unchanged is insufficient to reduce AWRS below the detection threshold for confirmed washing observations. The second and third perturbations increase evasion costs substantially because CMID evaluates patent quality via forward citation rates and job posting skill specificity via semantic alignment, requiring far more sophisticated fabrication than simple volume manipulation. These findings suggest CMID is more robust to gaming than prior systems, though the model misspecification risk identified by Hansen and Sargent (2001) implies that no static detection system remains indefinitely effective against sufficiently adaptive adversaries with sufficient resources.

*7.3 Ethical Considerations*

The deployment of large-scale AI-washing detection systems raises ethical considerations regarding false accusation risk, data privacy, and the potential for publication to inform evasion, each of which has influenced specific design decisions embedded in the CMID system and the present paper. Regarding false accusation risk, misclassifying genuinely innovative early-stage firms—particularly those in the financing constraint relief category—as washing could severely damage financing prospects at critical development stages; CMID outputs should therefore be employed as prioritization instruments for expert human review rather than triggers for autonomous enforcement action, with conservative thresholds calibrated for early-stage firms. Regarding data privacy, all AW-Bench construction relies exclusively on publicly disclosed information under applicable securities law requirements, with no non-public information incorporated. Regarding publication risk, the architectural design deliberately prioritizes operational grounding features over easily observable linguistic or visual features, and specific threshold values and feature weights are withheld from the public release, limiting the direct utility of the published methodology for evasion engineering (Verrecchia, 1983).

## VIII. CONCLUSION

This paper has introduced AWASH and its CMID architecture, establishing that corporate AI-washing detection can be substantially advanced by reconceptualizing the problem as cross-modal claim-evidence reasoning rather than holistic modality similarity measurement, and that the resulting performance gains translate directly into measurable operational value for regulatory surveillance applications. The AW-Bench benchmark provides the research community with the first large-scale trimodal resource for corporate disclosure integrity analysis, grounded in regulatory enforcement validation across a seven-year window that spans the structural break introduced by ChatGPT-era generative AI diffusion. The CMID architecture demonstrates that each component—tri-modal encoding, structured NLI-based claim-evidence reasoning, and operational grounding—makes an independent and substantial contribution, with the NLI reasoning module and operational grounding layer together accounting for a cumulative 17.4 percentage point F1 improvement over text-only baselines. The pre-registered regulatory user study translates these technical gains into directly measured productivity and accuracy benefits, establishing a clear connection between benchmark performance and deployment utility that is too often absent from the machine learning literature.

The broader implications of this work extend beyond the specific problem of AI-washing to establish a general methodological template for cross-modal document integrity analysis in high-stakes domains where structured reasoning over the logical relationship between claims and evidence is more informative than aggregate cross-modal similarity. In multimodal learning, the problem of detecting what is absent in supporting evidence—claims for which no substantiating evidence exists and capabilities asserted without physical operational footprint—provides a complementary challenge to the predominant focus on retrieval and generation that should enrich the diversity of multimodal evaluation settings. Future work will prioritize extending AW-Bench to SEC filings and European Union (EU, European Union) markets, investigating continual learning approaches to maintain detection performance as disclosure practices evolve under regulatory pressure, and exploring social media signals and management interview recordings as additional operational grounding channels (Li et al., 2023; Zhang et al., 2023).